# Approximate mechanism for measuring Stability of Internet Link in aggregated Internet Pipe

Vipin M, Mohamed Imran K R
{vipintm, mohamed.imran}@au-kbc.org
AU-KBC Research Centre, MIT Campus of Anna University,
Chennai, TN - INDIA

## Abstract

In this article we propose a method for measuring internet connection stability which is fast and has negligible overhead for the process of its complexity. This method finds a relative value for representing the stability of internet connections and can also be extended for aggregated internet connections. The method is documented with help of a real time implementation and results are shared. This proposed measurement scheme uses HTTP GET method for each connection(s). The normalized responses to identified sites like gateways of ISP's, google.com etc are used for calculating current link stability. The novelty of the approach is that historic values are used to calculate overall link stability. In this discussion, we also document a method to use the calculated values as a dynamic threshold metric. This is used in routing decisions and for load-balancing each of the connections in an aggregated bandwidth pipe. This scheme is a very popular practice in aggregated internet connections.

**Keyword: Internet measurement, routing, load balancing, failover**

## Introduction

Measuring link stability of individual internet connections in an aggregated pipe is highly recommended in any organization network. In an aggregated network, gateway handles all out-going traffic through multiple Internet connections provisions for load balancing and fail over.

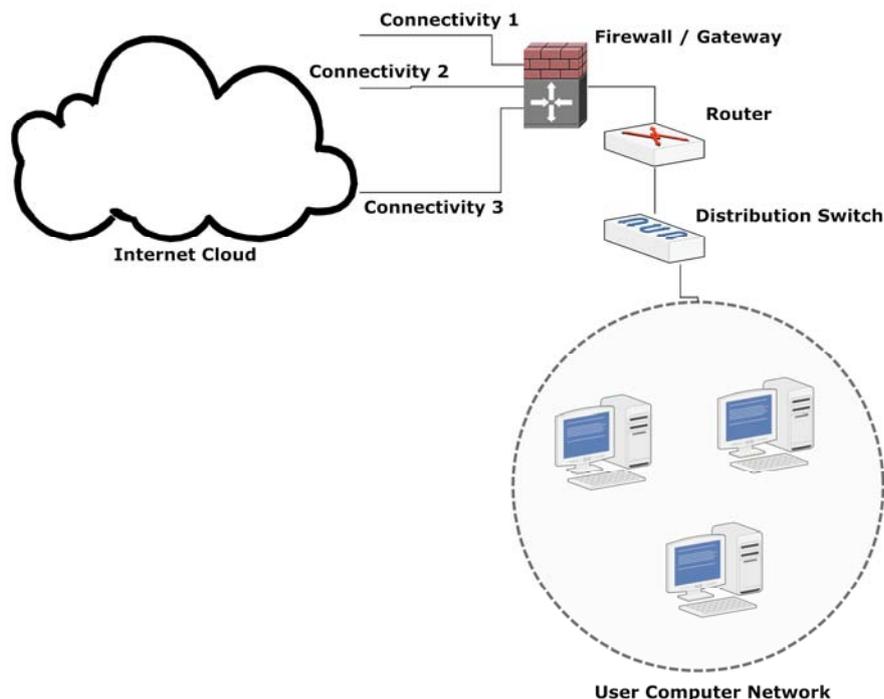

Figure 1

The out-going Internet connections can be from a single point or from multiple points. The stability measurement can give idea about per line stability/usability for a given observation time. This value can indicate the probability of that particular line to be usable in the future period with by referencing the historic values.

A simplified edge network for a typical organization is shown in Figure 1. The firewall or gateway router handles the connectivity to multiple internet connections. This individual line can be differentiated on bandwidth capacity, pricing and provider etc. These metrics are important in an optimized route selection algorithm for a better user/service experience. The ISP's provide service agreements on bandwidth and performance guarantees up to their edge routers. The user experience is not limited to the local sites and local services, so measuring and accounting performance with respect to world-wide access is required. As a result, we also consider stability and availability of individual connections as important for an optimal user /service experience.

The line stability information is suitable as a metric in routing decisions i.e. line selection for traffic and this can be achieve efficiently and easy. The measurement over a long period of time can give the complete picture of

- Line cost vs. fairness
- Contribution of stability from individual connections to the overall aggregated connection stability

## Scenario of Measurement

The most common method of checking link status is by using ping (ICMP). This method provides information like link status total round trip time to the specific server which can be defined to be reflective of world-wide access. This method is unable to determine the absolute link details of your Internet Service Provider (ISP) and the user experience as these responses from ICMP are only about layer-3 diagnostic information.

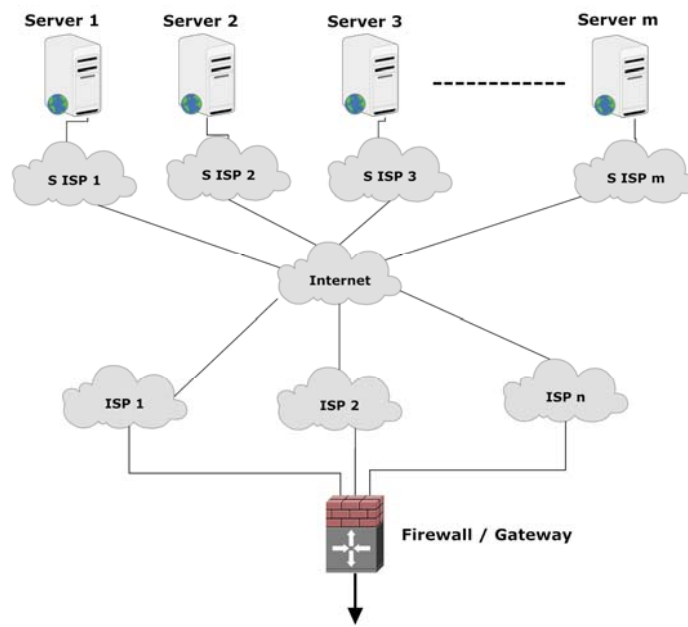

Figure 2

To measure the user experience, the employed method should mimic the same activity a normal user undertakes which is typically a few connections to multiple servers around the globe. These tests find the parameters for each link to different servers. As a base for our analysis, we need to identify a list of servers which are commonly used and have excellent stable responses like Google, Yahoo etc.

The Firewall/Gateway (F/G) is connected to the internet through multiple ISPs. Each ISP (ISP 1 – n) can have different routes of connectivity (depending on their infrastructure) to all other ISPs across world and they in turn, connect to different servers (Server 1 – m). For example, the average fetch time across the servers 1 – m for any ISP which is connected to F/G gives a measure of that ISP's current link status. The conditions are that the servers should be selected across trans-continental links and different ISP providers.

Before going to details of measurement system some terminologies are explained.

## Definitions

**Status:** This is a on / off status of the link and is represented as 0 or 1. If line is alive, it is 1 and for dead connection 0, it does not include any measure of connectivity. This is measured using a HTTP GET request to a stable website in server 1 – m.

**Tick:** This is a relative measuring of the link between two limits. This also measured using HTTP GET request to stable servers 1 – m and the values will be between 0 – m. Example if there are 5 selected servers then, the line is on best case 5 and out-of connection 0, and all other are mapped between these values.

**Stable Servers (websites):** These are the sites which give consistent performance on all links. E.g.: google.com, yahoo.com etc. Normally m is selected as 10 so there will be 10 different stable servers.

## Variable

**Stability Index (S):** The representative value of stability or the probability of line to be stable and (S * 100) is the percentage of stability.

**Line Index or Number (i):** This is a number which indicate the line in the case of multiple internet connections.

**Iteration Number (j):** Calculating the S is continues process, j represent the Iteration count of the process.

**Number of Lines (n):** This is the maximum value of i. And i = 1 to n

**Number of History to consider (k):** Number of history values to consider in the calculation of Historical Status.

**Line Status (L):** This is 1 or 0.

**Ticks (T):** This is a value between 0 and m.

**Ticks per iteration (m):** This is the maximum value map for the Ticks.

**No of History to calculate consistence value (z):** No of last history to use to calculate the constancy effect contributed because of changes in states of lines.

**Iteration consistence (R):** Constancy effect contributed because of changes in states of all lines in j[th] iteration.

**Consistency value (C):** The constancy effect contributed because of changes in states of all lines for last z history states.

## Measurement Method

In this measurement method the most important is calculating the Tics (T) for each link. This tick is calculated by doing HTTP GET request to 1 - m servers. On successful retrieval of page it is considers as 1 and 0 otherwise. The status of line (L) calculated form the T as;

$$L_{i,j} = \begin{cases} 0 & if \quad T_{ij} \leq 0 \\ 1 & if \quad T_{ij} > 0 \end{cases} \tag{1}$$

This Line status $L_{i,j}$ represent is the line i is alive or not in j[th] iteration. This value is used as a mask in the calculation of line stability.

The method is developed over a long window of observation of different factors in multiple internet lines. These lines are differentiated by bandwidth, provider and type of connectivity. There are two values are to be calculated

1. Link Stability Index per line
2. Total Internet pipe stability Index

**Link Stability Index per line**

The line stability for any line can calculate from four factors, and they are;

1. Line current status
2. Line last known Tick
3. Historical Ticks
4. Inherited effect from other lines

The line stability is given as;

$$S_{i,j} = \frac{1}{z \times m^2} \times L_{i,j} \times T_{i,j-1} \times H_{i,j} \times C_j \tag{2}$$

1) Line current status ($L_{i,j}$)

This is a flag for the current state and used to nullify the index value to zero when current status is out-of-connection.

2) Line last known status ($T_{i,j-1}$)

Last known status have highest impact on current and future status.

3) Historical status (H$_{i,j}$)

It considers the minimum from the history values and gives the verse state most probable to occur.

$$H_{i,j} = \min\left[T_{i,p}\right]_{p=j}^{j-k} \qquad (3)$$

4) Inherited effect from other lines (C$_{i,j}$)

This is a state wise count of all lines as an addictive maximum line state change. The inherited effect from other lines is calculated as the summation of iteration consistence of last z histories.

$$C_j = \sum_{r=j}^{j-z} R_r \qquad (4)$$

The iteration consistence or the consistence effect over all lines is found checking the state change between the current state and last known state.

$$R_r = \begin{cases} 1 & for \quad T_{i,j} = T_{i,j-1} \quad \forall i \\ 0 & other \quad wise \end{cases} \qquad (5)$$

The maximum possible value $z \times m^2$ is used as factorization to get the probability.

**Total Internet pipe stability Index**

The Internet pipe stability Index (IS) is calculated from the following factors

1. Total last known Tick All lines
2. Total of Link Historical status
3. Inherited effect from all lines

And it is given as;

$$IS_j = \frac{1}{z \times m \times n^2} \times \sum_{i=1}^{n} L_{ij} \times \sum_{i=1}^{n} H_{i,j} \times C_j \qquad (6)$$

The maximum possible value $z \times m \times n^2$ is used as factorization to get the probability.

# Implementation

This is implemented using Perl and Bash scripting and general Linux OS utilities [1]. The implementation was tested with three links which included two broadband and a leased line. The main script runs as an infinite loop with a sleeping time interval. In every loop it carry out three tasks they are;

1. Determining Ticks
2. Calculating the Stability Index
3. Routing and traffic class modification

### 1. Determining Ticks

There will be a list of m servers; in our implementation 10 sites are used. In every iteration j for every line i a HTTP GET is generated using wget [2] and the return value is checked with in a time out, on success it consider as 1 and 0 otherwise. So the m value vary from 0 – 10. At the same time the $L_{i,j}$ also found as in equation (1).

### 2. Calculating the Stability Index

This includes calculation of Internet Pipe stability index and each line stability index calculation. To do so it calculate $H_{i,j}$ first as in equation (3). In the initial state all the history values are considers as m in our case 10. Then it calculate $C_{i,j}$ by using equation (4), (5). For these calculation we used k=z=10. And $H_{i,j}$ calculation is done for all the n lines. And the $S_{ij}$ $IS_j$ is calculated.

### 3. Routing and traffic class modification

Based on the $IS_j$ and $S_{ij}$ we can optimize a lot of parameters for optimal performance. One of them is per link dynamic loading (based on the line stability). In the initial state the weight to each line is set based on the bandwidth (Bw) factor and it is set in the weighted round robin. And bandwidth factor given as;

$$Bwf_i = \left[ \frac{Bw_i}{\sum_{i=1}^{n} Bw_i} \right]^{\uparrow} \tag{7}$$

In each measurement iteration the weight can updated based on the $S_{ij}$ of each line. The dynamic routing weight can be obtained in different way; the method we followed is;

$$Rw_{ij} = \begin{cases} Bwf_i & if & S_{ij} \geq 0.95 \\ \dfrac{Bwf_i}{2} & if & 0.90 \leq S_{ij} < 0.95 \\ \dfrac{Bwf_i}{3} & if & 0.0 < S_{ij} < 0.90 \\ 0 & if & S_{ij} = 0 \end{cases} \tag{8}$$

By changing the routing weight dynamically the average packet sent to each line can be optimized depending on the derived capabilities based on our observations. In case any link is performing poorly because of the ISPs external connectivity problem or internal routing, delays will get affected in the $S_{ij}$ and thus the weight for that line will get reduced. This modification in weight of round robin will assign better links for more traffic and thus the user-experience is better.

Plot of $S_{ij}$ with respect to time for long period give a good understanding of the link. This plot does not represent bandwidth but represent the stability of each line and how long or how good the user experience will vary on this line.

Link failover decision can be made by watching the S$_{ij}$. Any time $S_{ij} \leq 0$ or the performance is not up to the minimum standard then that line can be removed from the routing path.

IS$_j$ can be used to decide on critical connection establishment like VPN etc. Example: It can be used for taking decision to accept any VPN connection in a specific line at any given time. In the case of multiple links, this allows us to determine which link is more stable based on the history, to accept a VPN like connection.

Like this using the IS$_j$ and S$_{ij}$ in the F/G level far more dynamic optimization can be achieved.

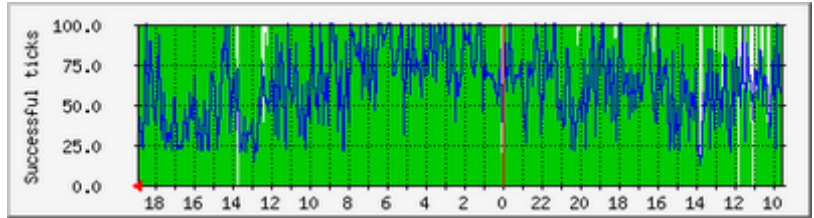
Figure 3

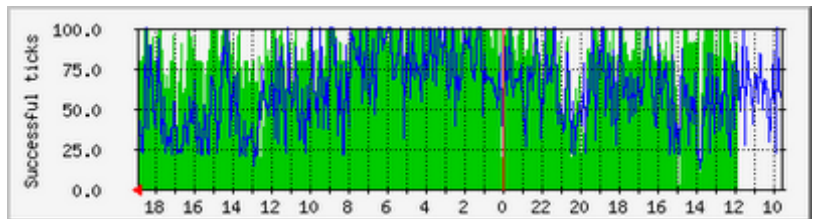
Figure 4

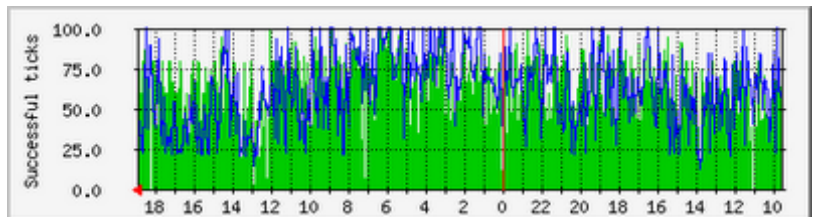
Figure 5

A three line stability percentage ( $S_{ij} \times 100$ ) Internet pipe stability percentage is shown in the Figure 3, 4 and 5. Figure 3 is Line-1, Figure 4 is Line-2 and Figure 5 is Line-3. In the figures the green line is the Line Stability and blue lines are Internet Stability. From the above three graphs the Line-1 is stable and the instability in the total Internet Pipe is contributed from the other two lines because the blue line is enveloped to those graphs. At this time S$_{ij}$ of Line-1 is 99% and the other two are less than 90%.

## Conclusion

In this article, we presented a measurement method to determine the stability of the internet links. Using our implementation as a reference, we found this method results in optimal usage of multiple internet links and results in better user experience.